\documentclass[conference]{IEEEtran}
\IEEEoverridecommandlockouts
\usepackage{cite}
\usepackage[numbers]{natbib}
\usepackage{amsmath,amssymb,amsfonts}
\usepackage{algorithmic}
\usepackage{graphicx}
\usepackage{textcomp}
\usepackage{xcolor}
\usepackage{svg}
\usepackage{tabularx}
\usepackage{booktabs}
\usepackage{subcaption}
\usepackage{multirow}
\usepackage{makecell}
\usepackage{cleveref}
\usepackage[inline]{enumitem}
\usepackage[]{authblk}

\newcolumntype{C}{>{\centering\arraybackslash}X}
\newcolumntype{R}{>{\raggedleft\arraybackslash}X}
\newcolumntype{L}{>{\raggedright\arraybackslash}X}

\usepackage{glossaries}
\newacronym{rssi}{RSSI}{Received Signal Strength Indicator}
\newacronym{poa}{PoA}{Phase-of-Arrival}
\newacronym{uwb}{UWB}{Ultra Wideband}

\usepackage{siunitx}
\DeclareSIUnit{\decibelm}{dBm}
\DeclareSIUnit{\decibeli}{dBi}
\DeclareSIUnit{\farad}{F}

\def\BibTeX{{\rm B\kern-.05em{\sc i\kern-.025em b}\kern-.08em
 T\kern-.1667em\lower.7ex\hbox{E}\kern-.125emX}}
\begin{document}

\newcommand{\bert}[1]{{\color{blue}[BERT: #1]}}
\newcommand{\liesbet}[1]{{\color{red}[LIESBET: #1]}}
\newcommand {\chesney}[1]{{\color{orange}[CHES: #1]}}

\title{Energy-Neutral Devices: \\ Can Hybrid RF-Acoustic Signals Point Them Out?

\thanks{This work is accepted by IEEE Asilomar 2020.\newline
This work was partially funded by the FWO project on location-locked cryptographic solutions for energy constrained devices under grant agreement G0D3819N}
}

\author[]{Bert Cox}
\author[]{Chesney Buyle}
\author[]{Liesbet Van der Perre}
\author[]{Lieven De Strycker}
\affil[]{Department of Electrical Engineering, KU Leuven, Belgium}

\maketitle

\begin{abstract}
We present a hybrid signaling approach to position energy-neutral devices. Our method combines the best of two worlds: instant RF signals for both communication and energy transfer, and slower propagating acoustic waves for accurate distance measurements. We introduce advanced energy loading of an ’E-buffer’ through beamformed RF signals. The scientific contribution of our approach is twofold, advancing both the positioning performance and the energy harvesting efficiency. On the one hand, we overcome current distance limitations in RF backscattering-based indoor localization. On the other hand, it enhances the energy harvesting by using the calculated position and beamforming a higher amount of directed RF energy into the mobile node. We provide a functional architecture for both sides of the system and present a proof-of-concept. Practical measurements in representative use cases show an update rate of 10 positions per hour within the regulatory constraints in the 868MHz band for distances up to \SI{4.5}{\meter}. An energy and power model are drawn up to provide insights into the performance trade-offs. 
\end{abstract}

\begin{IEEEkeywords}
Acoustic Sensors, Hybrid Signaling, Distance Measurement, Low-Power Electronics, Beamforming
\end{IEEEkeywords}

\section{Introduction}
Location-based services have recently received significant interest, driven by numerous applications in domains such as asset monitoring, access control, indoor navigation and tracking~\cite{basiri2017indoor}. Consequently, there is a strong need for reliable and accurate real-time indoor identification and localization. While a wide variety of sensing technologies and methods have been proposed, there is still no indoor positioning solution that can check all the boxes. According to Zafari et al. a localization system should be cost efficient, energy efficient, have a wide reception range, high localization accuracy, low latency and high scalability in order to be wide-scale adopted~\cite{8692423}. 

A sought-after feature in many applications is fully-passive operation of the tracking device. As no local power source such as a battery is required, this will reduce maintenance and device costs while increasing flexibility and lifetime. For positioning of fully-passive devices, RFID has become the main technology of choice.  However, the accuracy of RFID localization systems based on \gls{rssi} is limited (order of meters) and is strongly affected by multipath effects and tag orientation~\cite{5451187}\cite{5504205}. \gls{poa} methods typically show a better localization accuracy, but currently suffer from two main issues: ambiguity due to phase periodicity and phase offset~\cite{6469211}.  

Many of the most promising technologies that meet many quality metrics for indoor positioning are \gls{uwb}-signaling based. In \gls{uwb} localization, the system usually relies on the time of flight of sub-nanosecond pulses to calculate the distance between the transmitter and target~\cite{6310018}. Since \gls{uwb} uses a very large frequency bandwidth ($>$\SI{500}{\MHz}), a high resolution in time and thus distance can be obtained, resulting in a positioning accuracy in the order of tens of centimeters~\cite{6887301}. In addition, the high bandwidth ensures great resistance against multipath fading and consequently makes \gls{uwb} highly attractive for indoor environments. In order to coexist with narrowband system architectures, the spectral power density in \gls{uwb} is limited to \SI[per-mode=symbol]{-41.3}{\decibelm\per\MHz} by regulations. Unfortunately, this is insufficient to power \gls{uwb} tags at useful distances~\cite{6310018}\cite{6205975}. 


Extensive research has been conducted on acoustic localisation systems. As acoustic signals have a relative low propagation speed (343 m/s), centimeter positioning accuracy can be achieved based on time of flight measurements~\cite{medina2013ultrasound}. A disadvantage in contrast to UWB, however, is that acoustic signals are susceptible to multipath interference. In most acoustic localisation architectures, the acoustic channel is combined with a Radio Frequency (RF) channel to form a hybrid system. The RF link is used primarily for synchronisation for distance measurements, yet it can also serve as a communication channel. Although the energy efficiency of wireless acoustic sensor nodes has progressed significantly because of the developments in MEMS microphones and low power computing power, current hybrid RF-acoustic solutions have not achieved fully passive operation. Research has shown that the RF communication is the main limiting factor to achieve this~\cite{9078776}. 

In our work we exploit two main techniques in order to meet the power requirements of the hybrid RF-acoustic tag without batteries: RF backscatter and energy harvesting. In RF backscatter, the tag is able to transmit information by modulating the incident electromagnetic field from an external source through its antenna impedance, improving the energy efficiency for RF communication by multiple order of magnitudes~\cite{talla2017lora}. The main disadvantage, however, is that the reflected wave will experience a significant signal-to-noise ratio (SNR) drop because of the propagation distance that needs to be traveled twice. To cover the remaining power needs, the hybrid RF-acoustic tag can harvest energy from the environment. However, the power density of these ambient sources is too low in comparison to the energy requirements of the acoustic ranging system presented in~\cite{9078776} to serve as a useful energy source~\cite{6951347}. One way to overcome this problem is to steer a directional RF power beam towards the tag, significantly increasing the power density~\cite{Yedavalli2017beamWPT}\cite{choi2019}. 

In this paper, we present a hybrid RF-acoustic signaling architecture for the localisation of energy-neutral devices. We extend current low power RF-acoustic positioning systems with RF backscattering and directional wireless power transfer to enable full passive operation of a mobile tag. A functional system architecture is provided for both the beacon and passive tag. As a first contribution, this paper proposes the ranging and positioning strategy of a mobile tag based on chirp frequency demodulation. Secondly, the performance of a RF energy harvester is evaluated through simulation and practical measurements in a non-anechoic environment.

This paper is further organized as follows. First, we discuss the signaling strategy of the hybrid RF-acoustic ranging system. Next, we present a functional system architecture. In Section~\ref{sec:calcandsim}, the performance of the system is evaluated through energy calculations and simulations. The practical measurement setup and procedures are presented in Section~\ref{sec:measurementsetup}. Finally, the results (Section~\ref{sec:results}), conclusions and future steps are presented (Section~\ref{sec:Conc}). 



\section{The Hybrid RF-Acoustic System}

\subsection{Ranging and Positioning Strategy}
First, the one-dimensional ranging strategy is introduced. Based on this, the system can be extended intuitively for two- or three-dimensional positioning. An overview of the ranging strategy is shown in Fig.~\ref{fig:systemoverview} and is based on the chirp frequency demodulation technique presented in~\cite{9078776}. The system consists of two types of entities: a fixed beacon (B) and one or multiple mobile tags ($M_x$). The distance between the beacon and tag is obtained as follows:

\begin{figure}[]
 \centering
\includegraphics[width=0.45\textwidth, angle=0]{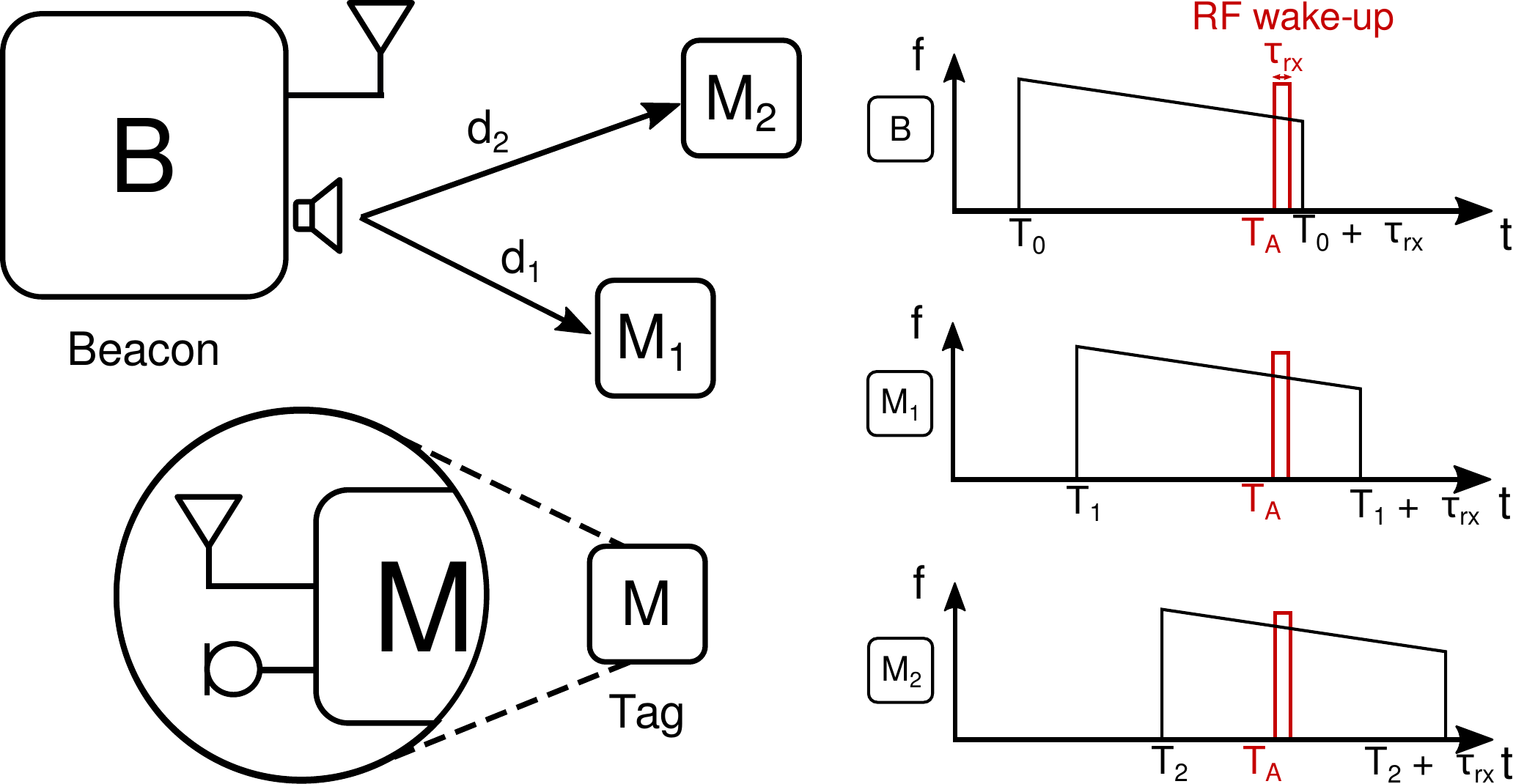}
\caption{Ranging strategy: a beacon periodically broadcasts an ultrasonic (US) chirp signal. Upon reception of a RF wake-up signal, all mobile tags capture a specific part of the US chirp.}
\label{fig:systemoverview}
\vspace{-0.4cm}
\end{figure}

\begin{enumerate}
    \item The beacon starts broadcasting an ultrasonic chirp of duration $\tau_{tx}$ at time $T_0$.
    \item At a given time $T_A \geq T_0$, the beacon sends out an RF wake-up signal. Upon reception, each tag samples the received acoustic signal for a short period of time $\tau_{rx}$. Since all tags wake up at nearly the same moment, every tag will receive a specific part of the ultrasonic chirp depending on their distance to B.
    \item Each tag sends back the collected audio data to the beacon via an RF link. 
    \item Ultimately, the time of flight (TOF), and thus distance, between both entities can be obtained through cross-correlation of the received audio part and the broadcasted audio signal. 
\end{enumerate}
The energy consumption of the tag can be kept low by limiting the acoustic receive window $\tau_{rx}$. However, this adversely impacts the accuracy of the cross-correlation and hence distance measurement. Consequently, there exists a trade-off between the accuracy of the ranging system and the tag's energy consumption. 
The system can be extended intuitively for two- or three-dimensional positioning by introducing at least two and three more beacons respectively. 


\subsection{Functional Architecture Passive Hybrid RF-Acoustic Tag}
\begin{figure}[]
 \centering
\includegraphics[width=0.4\textwidth, angle=0]{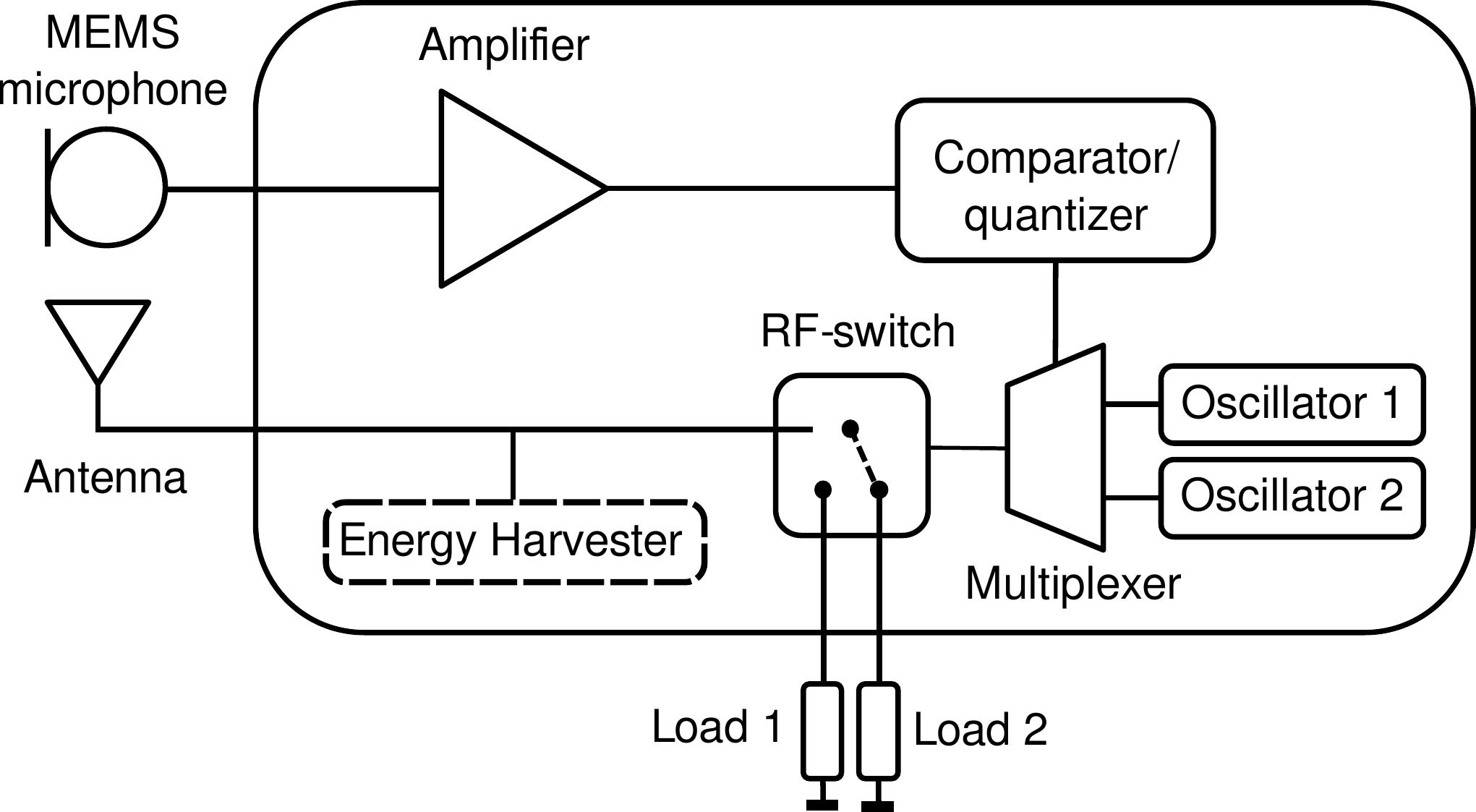}
\caption{Functional diagram of the hybrid RF-acoustic tag.}
\label{fig:tag}
\vspace{-0.4cm}
\end{figure}

 The system architecture of a hybrid RF-acoustic tag based on RF backscatter is presented here, further improving energy consumption with respect to the work presented in~\cite{9078776} which is based on an active radio transceiver. The functional diagram is shown in Fig.~\ref{fig:tag}. 


The ultrasonic chirp signal is picked up by a MEMS microphone, amplified through two amplifier stages and quantified via a comparator to a binary signal stream. This binary stream controls the output of a multiplexer, switching its output between two oscillator signals with a small mutual frequency shift. The output of the multiplexer is in turn connected to an RF switch. As the microphone signal alternates, the output of the comparator will switch between zero and one and the RF switch will be controlled by either one of the two oscillators. At the rate of the oscillator in control, the RF switch will connect one of the two loads to the tag's antenna. The loads must be chosen such that difference of the tag's radar cross section is maximized~\cite{6206613}.

During the receive window $\tau_{rx}$, the captured part of the chirp is modulated on the incident RF wave through Frequency Shift Keying (FSK). Ultimately, the reflected RF signal is sampled by the beacon and again, a cross-correlation can be performed to obtain the distance. 

\subsection{RF power beaming}
In order to enable fully passive operation, power is added to the tag through a directional RF power beam. An RF energy harvester is included in the design of the tag to capture the energy of the RF field (Fig.~\ref{fig:tag}). At the beacon's side, a phased antenna array is used to steer the directional RF beam towards the mobile tag. To this end, the beam angle can be calculated from the aforementioned localisation system.

The phased antenna array at the beacon helps to overcome two main limitations in Wireless Power Transfer (WPT): path loss and interference with communication systems. According to the Friis transmission formula, free-space path loss increases with the distance squared. Assuming a constant operating frequency and transmission power, the gain of the transmitter and receiver antenna can be increased to compensate for this loss. However, since the tag is mobile and a random orientation must be considered, the tag's antenna must be kept omnidirectional to harvest from any direction. Consequently, a phased antenna array is considered to increase the transmitter's gain and thus received power at the tag. It should be noted, however, that the transmission power and transmitter's gain are usually limited by international regulations. Furthermore, since the RF power beam is steered directional towards the mobile tag, interference issues with wireless data communication systems can be lowered. 

The full operation of the passive hybrid RF-acoustic positioning system consists of two phases: the pre-charge phase and the ranging phase. In the first phase, the energy buffer (E-buffer) of the tag is charged through the directional RF power beam. Initially, the beacon does not know its relative position to the tag. Hence, a RF beam sweep can be executed to overcome the initial charge. In the second phase, distance measurements are carried out and the position of the tag can be calculated. 

\section{Mathematical and simulation-based analysis}
\label{sec:calcandsim}
In this section, the performance of the system is evaluated through calculations and simulations in terms of the tag's charge time, i.e. the time it takes for the beacon to pre-charge the tag's E-buffer with sufficient energy to perform one ranging measurement. First, the energy and power requirements of the tag are calculated based on off-the-shelf components\footnote{Microphone: Knowles SPU0410, amplifiers: TI TLV341A, comparator: TI TLV3691, multiplexer: TI TS5A3160, oscillators: Touchstone TS3001, RF-switch: AD ADG904} and the system operation. From this, a suitable size for the tag's energy buffer is determined in accordance with the design guidelines of an off-the-shelf RF energy harvester. Finally, the charge time of the proposed system is simulated based on the Friis transmission formula. 

\subsection{Tag Energy and Power Requirements}

\begin{table}
	\centering
	\caption{Power consumption $P$ and turn-on time $t_{\text{on}}$ of the tag's functional components. } 
	\label{tab:powerturon}
	\renewcommand{\arraystretch}{}
	\begin{tabularx}{\linewidth}{@{}lCCCCCC}
		\toprule
    		& Mic. & Amp. (2x) & Comp. & Mux & Osc. (2x) & RF-switch \\
		\midrule                                   
            $\mathbf{P\,[\si{\micro W]}}$ & 216 & 252 & 0.135 & 0.009 & 10.08 & 1.8 \\
            $\mathbf{t_{\text{on}}}\,[\si{\milli\second}]$ & 50 & 0.005 & 0.2 & 0.000013 & 1.75 & 0.0000215 \\
		\bottomrule
	\end{tabularx}
	\vspace{-0.4cm}
\end{table}

The power consumption and turn-on time of each element in the tag's functional diagram is shown in Table~\ref{tab:powerturon}. It shows that the ultrasonic MEMS microphone has a relatively high power consumption and large turn-on time, while the amplifier has an even higher power consumption, but negligible turn-on time. Instead of powering up all components simultaneously for \SI{50}{\milli\second}, it follows from Table~\ref{tab:powerturon} that a split start-up, in which the MEMS microphone is turned on in the first \SI{50}{\milli\second} and the other components in the final \SI{1.75}{\milli\second}, reduces the turn-on energy consumption of the tag by more than \SI{50}{\percent}. In accordance with the ranging strategy proposed in~\cite{9078776}, the tag is assumed to be powered for \SI{1}{\milli\second} after start-up. The total energy consumption $E_{tag}$ to perform one ranging measurement using a split start-up is calculated to be \SI{11.7}{\micro\joule}, based on the component's technical datasheet information in Table~\ref{tab:powerturon}. 

\subsection{RF Energy Harvester Configuration}
The energy harvester's E-buffer must be sized carefully. An oversized energy buffer may require a long time to reach the targeted voltage, while an undersized energy buffer contains too little energy to perform the system operations. For this application, the AEM40940 RF energy harvester~\cite{AEM4940} from E-peas is chosen. Since only a small amount of energy is required ($E_{tag} = \SI{11.7}{\micro\joule}$), a capacitor is used as the tag's energy buffer. 

The size of the capacitor is calculated based on two harvester specific voltage levels, $V_{chrdy}$ and $V_{ovdis}$, and the harvester's LDO efficiency. When an RF power between \SI{-19.5}{\decibelm} and \SI{10}{\decibelm} is present at the input of the energy harvester, it starts charging the capacitor. The moment the voltage over the capacitor becomes equal to $V_{chrdy}$, the harvester is ready to deliver power to the tag's components. A voltage regulator (LDO) assures that the capacitor voltage is converted into a stable and constant \SI{1.8}{\volt} as the tag's power source. During the system operations, the capacitor voltage will gradually decrease to $V_{ovdis}$. At this voltage, the energy harvester will cut off the tag's power supply in order to prevent overdischarge of the energy buffer. Consequently, the minimum capacitor's size $C_{min}$ can be calculated as follows:

\begin{equation}
\label{eq:cmin}
    C_{min} \geq \frac{2E_{tag}}{\eta_{LDO} (V_{chrdy}^2 - V_{ovdis}^2) },
\end{equation} 

where $E_{tag}$ is the energy required to perform one ranging measurement and $\eta_{LDO}$ is the voltage regulator's conversion efficiency. 

In the case of a single cell capacitor, $V_{chrdy} = \SI{2.30}{\volt}$ and $V_{ovdis} = \SI{2.20}{\volt}$. The voltage regulator efficiency $\eta_{LDO}$ depends on the capacitor's voltage and the output voltage. For a output voltage of \SI{1.8}{\volt}, the worst case efficiency occurs when the capacitor voltage equals $V_{chrdy}$. In this case, $\eta_{LDO} = 0.77$. From Eq.~\ref{eq:cmin} follows that $ C_{min} \geq \SI{67.8}{\micro\farad}$. Considering a commercially available capacity of $C = \SI{68}{\micro\farad}$, a total energy $E_{cap} = \frac{C(V_{chrdy}^2 - V_{ovdis}^2)}{2} = \SI{15.3}{\micro\joule}$ must be harvested.   

\subsection{Performance Simulation}
\begin{figure}[]
 \centering
\includegraphics[width=0.4\textwidth, angle=0]{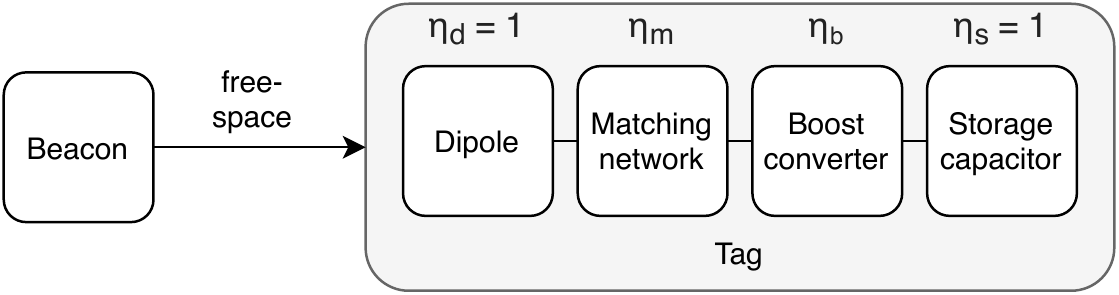}
\caption{Overview of the simulation process with the different efficiencies.}
\label{fig:simulation}
\vspace{-0.4cm}
\end{figure}

An overview of the simulation process is shown in Fig.~\ref{fig:simulation}. The beacon and tag are considered to be in a free-space environment. 

The beacon transmits a sinuso\"idal continuous wave (CW) at a frequency of \SI{869.5}{\mega\hertz}. According to the ETSI regulations, the maximum Effective Isotropic Radiated Power (EIRP) in this frequency band is limited to \SI{27}{\decibelm} with a maximum duty cycle of 10\%. Consequently, the transmission power $P_t$ and transmitter $G_t$ are considered to be \SI{27}{\decibelm} and \SI{0}{\decibeli} respectively. At the tag, a dipole antenna with gain $G_r = \SI{2.15}{\decibeli}$ captures the incident RF power beam. The antenna efficiency $\eta_d$ is assumed to be \SI{100}{\percent}. The antenna signal passes through a matching network with efficiency $\eta_{m}$ and is then boosted with efficiency $\eta_{b}$ by the AEM40940 RF energy harvester. Finally, the boost converter delivers a current to the storage element. No losses ($\eta_s = 1$) are assumed during this transfer. Based on the Friis transmission formula and the energy harvesting efficiency, the power delivered to the capacitor $P_{harvest}$ in function of the distance $d$ is calculated as follows:

\begin{equation}
P_{harvest} = \eta_{m}\eta_{b}\frac{G_t G_r P_t \lambda^2}{(4 \pi d)^2 } 
\end{equation}

The efficiency $\eta = \eta_{m}\eta_{b}$ is provided by the E-peas AEM40940 technical datasheet~\cite{AEM4940}. Ultimately, the charge time $\Delta t_{charge} = E_{cap}/P_{harvest}$ is calculated for two scenarios.
\begin{enumerate}
    \item Initial charge time $t_{initial}$ = time it takes for the energy harvester to charge the capacitor from \SI{0}{V} to $V_{chrdy} = \SI{2.3}{V}$. This scenario occurs when the tag has not been powered for a long time.
    \item Update charge time $t_{update}$ = time it takes for the energy harvester to charge the capacitor from $V_{ovdis} = \SI{2.2}{V}$ to $V_{chrdy} = \SI{2.3}{\volt}$. This scenario occurs when a ranging measurement has recently taken place. 
\end{enumerate}

\begin{figure}[]
 \centering
\includegraphics[width=0.3\textwidth, angle=0, trim = 3.5cm 8.5cm 3.6cm 10cm]{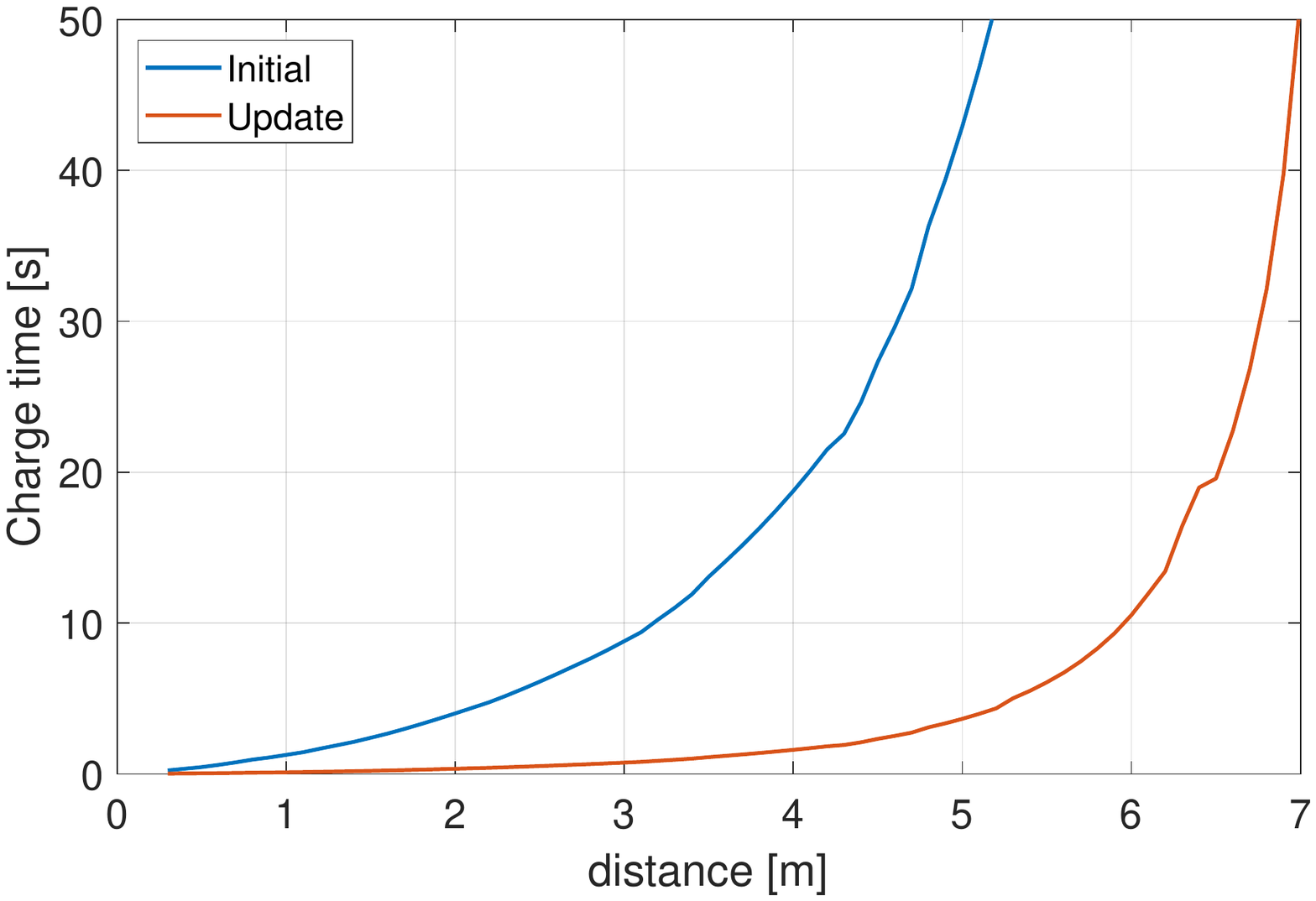}
\caption{Simulated initial and update charge time in function of distance.}
\label{fig:simulated_chargetime}
\vspace{-0.4cm}
\end{figure}
 
Fig.~\ref{fig:simulated_chargetime} shows the simulated initial and update charge time in function of the distance. The charge time remains below \SI{30}{\second} at a distance of \SI{4.5}{\meter} for the initial charge case and takes around \SI{10}{\second} at a distance of \SI{6}{\meter} for the update charge case. At distances larger than \SI{4.5}{\meter} and \SI{6.5}{\meter} for respectively the initial and update charge cases, the charge times increase significantly. This can be explained by the decreasing efficiency $\eta$ of the AEM40940 energy harvester at decreasing input power. 
 

\section{Measurement setup}
\label{sec:measurementsetup}
\begin{figure}[]
 \centering
\includegraphics[width=0.48\textwidth]{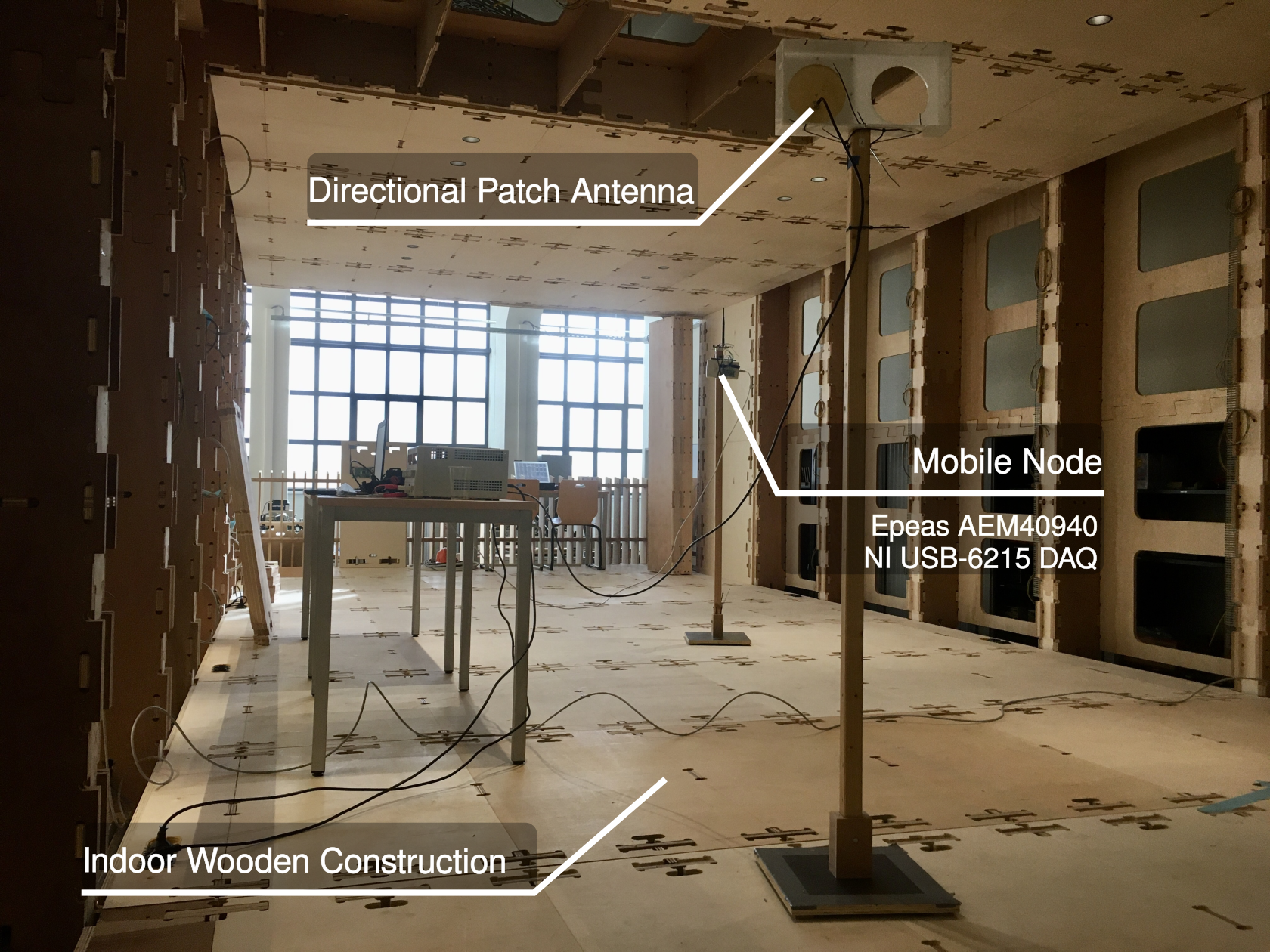}
\caption{Measurement setup: transmit beacon with directional patch antenna (front pole) and mobile node with the $Epeas AEM40940$ Energy harvester (back pole).}
\label{fig:measurementsetup}
\vspace{-0.4cm}
\end{figure}

The performance of the energy harvester part of the tag was evaluated through practical measurements. A transmit beacon and mobile node were set up in an indoor wood construction, shown in Fig.~\ref{fig:measurementsetup}. Both entities were attached to a wooden pole at a height of \SI{1.65}{\meter}. A sinuso\"idal continuous wave (CW) at a frequency of \SI{869.5}{\mega\hertz} was transmitted by the beacon through a directional patch antenna with gain $G_t$ \SI{6.15}{\decibeli}. The transmit power of the beacon was adjusted to the maximum EIRP of \SI{27}{\decibelm}. At the mobile node, a dipole antenna with gain $G_r$ of \SI{2.15}{\decibeli} was connected to the Epeas AEM40940 energy harvester. A single cell capacitor of \SI{68}{\micro\farad} was used as the energy harvester's E-buffer. 

The charge time of the energy harvester was measured for the same two scenarios as described in the simulation section, i.e. the initial charge time $t_{initial}$ and the update charge time $t_{update}$. The charge time measurements were conducted along two arbitrary straight paths inside the construction. The transmitter patch antenna was directed towards the mobile node in the direction of maximal gain at all times. The distance between the beacon and mobile node was increased in steps of \SI{0.5}{\meter} starting at \SI{1}{\meter} and ending at \SI{7}{\meter}. For each distance, respectively 25 and 100 measurements were performed for the initial and update charge time. To partially measure the influence of interference (constructive or destructive), RF absorbing panels were added around the receiving antenna in one of the measurement sequences.

\section{Results}
\label{sec:results}
\begin{figure}[]
 \centering
\includegraphics[width=0.45\textwidth, angle=0]{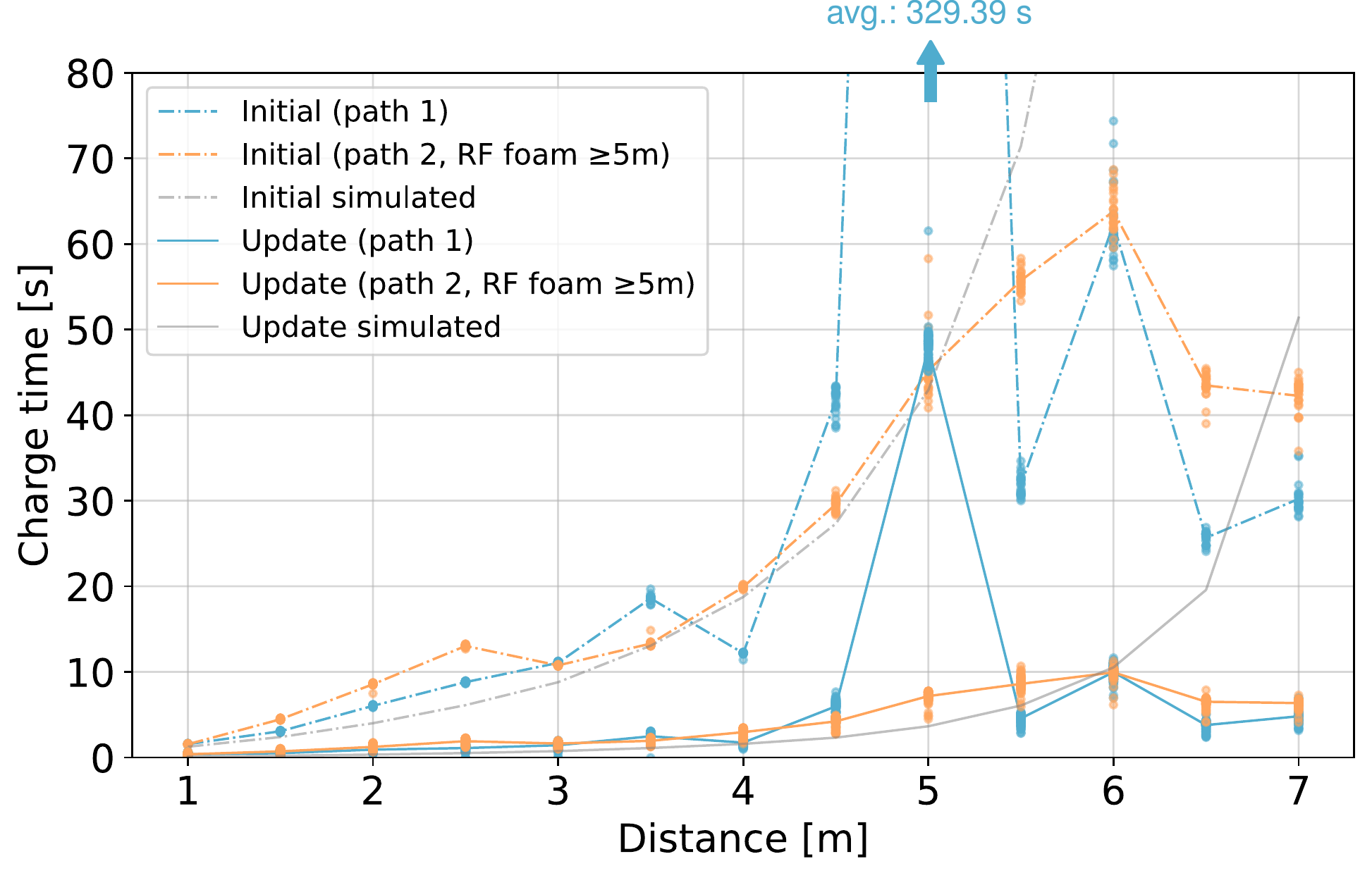}
\caption{Initial (dashed line) and update (solid line) charge time measurements at different distances.}
\label{fig:initialandupdatechargetime}
\vspace{-0.4cm}
\end{figure}

Fig.~\ref{fig:initialandupdatechargetime} shows the results of the initial and update charge time measurements. RF absorbing panels were used in the case of path 2 at distances $\geq$ \SI{5}{\meter}. For the initial charge time case, the measurements follow the simulated values up to \SI{4.5}{\meter}. The destructive interference that occurs at \SI{5}{\meter} along path 1 initially also appeared in the case of path 2, but disappeared when RF absorbers were placed. For distances larger than \SI{5}{\meter}, better charge times than simulated are obtained for both scenarios. Future measurements in an anechoic room will provide an in-depth analysis of the occurring interference. Practically, these numbers mean that we can achieve a positioning update rate of 1 location per minute for all distances and even up to 6 position updates for distances smaller than \SI{2}{\meter}. However, when we incorporate the ETSI duty cycle regulations, the update rate drops drastically to around 10 position updates per hour at a distance of \SI{4.5}{\meter}. The initial charge time measurements represent the worst case scenario where the capacitor is fully discharged. However, as the update rate increases, so does the amount of power left in the capacitor. Hence, the update charge time in Fig.~\ref{fig:initialandupdatechargetime} resembles the initial charge time, following the theoretical values at lower distances. The charge time remains lower than \SI{10}{\second}, resulting in a potential position update rate of 6 times per minute. Taking the duty cycle limitation into account, a position can be calculated every \SI{100}{\second} in the ideal scenario.

\section{Conclusions and Future Work}
\label{sec:Conc}
In the present study, we show that directional wireless power transfer in combination with RF backscattering enables the full passive positioning of a hybrid RF-acoustic mobile node. Through practical measurements in a non-anechoic environment, it was found that the initial positioning update rate is around 10 locations per hour at a distance of \SI{4.5}{\meter}. For the update charge scenario, this rate increases to 1 location update every \SI{100}{\second}. Applications can be found in positioning patient monitoring equipment in hospitals or indoor tracking of large parcels. Main limitations in the lower positioning estimate is the ETSI regulations where both the EIRP and duty cycle are limited. A more detailed investigation in dual band operation and polite spectrum access techniques should reveal if these limitations could be countered. The outliers and improved results at larger distances are likely due to destructive and constructive interference. Future research is needed to determine the cause of these deviations. It should be noted that in this research, the beamforming only targets a single mobile node. Further analyses should investigate how the system can be extended to 3D positioning for a large number of mobile nodes. 


{\footnotesize
\bibliography{bronnen.bib}}%

\begin{thebibliography}{16}
\providecommand{\natexlab}[1]{#1}
\providecommand{\url}[1]{#1}
\csname url@samestyle\endcsname
\providecommand{\newblock}{\relax}
\providecommand{\bibinfo}[2]{#2}
\providecommand{\BIBentrySTDinterwordspacing}{\spaceskip=0pt\relax}
\providecommand{\BIBentryALTinterwordstretchfactor}{4}
\providecommand{\BIBentryALTinterwordspacing}{\spaceskip=\fontdimen2\font plus
\BIBentryALTinterwordstretchfactor\fontdimen3\font minus
  \fontdimen4\font\relax}
\providecommand{\BIBforeignlanguage}[2]{{%
\expandafter\ifx\csname l@#1\endcsname\relax
\typeout{** WARNING: IEEEtranN.bst: No hyphenation pattern has been}%
\typeout{** loaded for the language `#1'. Using the pattern for}%
\typeout{** the default language instead.}%
\else
\language=\csname l@#1\endcsname
\fi
#2}}
\providecommand{\BIBdecl}{\relax}
\BIBdecl

\bibitem[Basiri et~al.(2017)Basiri, Lohan, Moore, Winstanley, Peltola, Hill,
  Amirian, and e~Silva]{basiri2017indoor}
A.~Basiri, E.~S. Lohan, T.~Moore, A.~Winstanley, P.~Peltola, C.~Hill,
  P.~Amirian, and P.~F. e~Silva, ``Indoor location based services challenges,
  requirements and usability of current solutions,'' \emph{Computer Science
  Review}, vol.~24, pp. 1--12, 2017.

\bibitem[{Zafari} et~al.(2019){Zafari}, {Gkelias}, and {Leung}]{8692423}
F.~{Zafari}, A.~{Gkelias}, and K.~K. {Leung}, ``A survey of indoor localization
  systems and technologies,'' \emph{IEEE Communications Surveys Tutorials},
  vol.~21, no.~3, pp. 2568--2599, 2019.

\bibitem[{Errington} et~al.(2010){Errington}, {Daku}, and {Prugger}]{5451187}
A.~F.~C. {Errington}, B.~L.~F. {Daku}, and A.~F. {Prugger}, ``Initial position
  estimation using {RFID} tags: A least-squares approach,'' \emph{IEEE Trans.
  on Instr. and Meas.}, vol.~59, no.~11, pp. 2863--2869, 2010.

\bibitem[{Saab} and {Nakad}(2011)]{5504205}
S.~S. {Saab} and Z.~S. {Nakad}, ``A standalone {RFID} indoor positioning system
  using passive tags,'' \emph{IEEE Transactions on Industrial Electronics},
  vol.~58, no.~5, pp. 1961--1970, 2011.

\bibitem[{DiGiampaolo} and {Martinelli}(2014)]{6469211}
E.~{DiGiampaolo} and F.~{Martinelli}, ``Mobile robot localization using the
  phase of passive {UHF} {RFID} signals,'' \emph{IEEE Transactions on
  Industrial Electronics}, vol.~61, no.~1, pp. 365--376, 2014.

\bibitem[{Cruz} et~al.(2013){Cruz}, {Costa}, and {Fernandes}]{6310018}
C.~C. {Cruz}, J.~R. {Costa}, and C.~A. {Fernandes}, ``Hybrid {UHF}/{UWB}
  antenna for passive indoor identification and localization systems,''
  \emph{IEEE Trans. on Ant. and Prop.}, vol.~61, no.~1, pp. 354--361, 2013.

\bibitem[{Zwirello} et~al.(2015){Zwirello}, {Schipper}, {Jalilvand}, and
  {Zwick}]{6887301}
L.~{Zwirello}, T.~{Schipper}, M.~{Jalilvand}, and T.~{Zwick}, ``Realization
  limits of impulse-based localization system for large-scale indoor
  applications,'' \emph{IEEE Transactions on Instrumentation and Measurement},
  vol.~64, no.~1, pp. 39--51, 2015.

\bibitem[{Cruz} et~al.(2012){Cruz}, {Costa}, and {Fernandes}]{6205975}
C.~C. {Cruz}, J.~R. {Costa}, and C.~A. {Fernandes}, ``Design of a passive tag
  for indoor localization,'' in \emph{2012 6th European Conference on Antennas
  and Propagation (EUCAP)}, 2012, pp. 2495--2499.

\bibitem[Medina et~al.(2013)Medina, Segura, and De~la
  Torre]{medina2013ultrasound}
C.~Medina, J.~C. Segura, and A.~De~la Torre, ``Ultrasound indoor positioning
  system based on a low-power wireless sensor network providing sub-centimeter
  accuracy,'' \emph{Sensors}, vol.~13, no.~3, pp. 3501--3526, 2013.

\bibitem[{Cox} et~al.(2020){Cox}, {Van der Perre}, and {De Strycker}]{9078776}
B.~{Cox}, L.~{Van der Perre}, and L.~{De Strycker}, ``Zero-crossing chirp
  frequency demodulation for ultra-low-energy precise hybrid {RF}-acoustic
  ranging of mobile nodes,'' \emph{IEEE Sensors Letters}, vol.~4, no.~5, pp.
  1--4, 2020.

\bibitem[Talla et~al.(2017)Talla, Hessar, Kellogg, Najafi, Smith, and
  Gollakota]{talla2017lora}
V.~Talla, M.~Hessar, B.~Kellogg, A.~Najafi, J.~R. Smith, and S.~Gollakota,
  ``Lora backscatter: Enabling the vision of ubiquitous connectivity,''
  \emph{Proceedings of the ACM on Interactive, Mobile, Wearable and Ubiquitous
  Technologies}, vol.~1, no.~3, pp. 1--24, 2017.

\bibitem[{Lu} et~al.(2015){Lu}, {Wang}, {Niyato}, {Kim}, and {Han}]{6951347}
X.~{Lu}, P.~{Wang}, D.~{Niyato}, D.~I. {Kim}, and Z.~{Han}, ``Wireless networks
  with {RF} energy harvesting: A contemporary survey,'' \emph{IEEE
  Communications Surveys Tutorials}, vol.~17, no.~2, pp. 757--789, 2015.

\bibitem[{Yedavalli} et~al.(2017){Yedavalli}, {Riihonen}, {Wang}, and
  {Rabaey}]{Yedavalli2017beamWPT}
P.~S. {Yedavalli}, T.~{Riihonen}, X.~{Wang}, and J.~M. {Rabaey}, ``Far-field rf
  wireless power transfer with blind adaptive beamforming for internet of
  things devices,'' \emph{IEEE Access}, vol.~5, pp. 1743--1752, 2017.

\bibitem[{Choi} et~al.(2019){Choi}, {Ginting}, {Aziz}, {Setiawan}, {Park},
  {Hwang}, {Kang}, {Chung}, and {Kim}]{choi2019}
K.~W. {Choi}, L.~{Ginting}, A.~A. {Aziz}, D.~{Setiawan}, J.~H. {Park}, S.~I.
  {Hwang}, D.~S. {Kang}, M.~Y. {Chung}, and D.~I. {Kim}, ``Toward realization
  of long-range wireless-powered sensor networks,'' \emph{IEEE Wireless
  Communications}, vol.~26, no.~4, pp. 184--192, 2019.

\bibitem[{Dimitriou} et~al.(2012){Dimitriou}, {Bletsas}, and
  {Sahalos}]{6206613}
A.~G. {Dimitriou}, A.~{Bletsas}, and J.~N. {Sahalos}, ``Practical
  considerations of {ASK} modulated passive tags,'' in \emph{2012 6th European
  Conference on Antennas and Propagation (EUCAP)}, 2012, pp. 3476--3480.

\bibitem[E-p(2018)]{AEM4940}
\BIBentryALTinterwordspacing
\emph{Highly-efficient, regulated dual-output, ambient energy manager for
  high-frequency RF input with optional primary battery}, E-peas, 2018, rev.
  1.1. [Online]. Available: \url{https://e-peas.com/product/aem40940/}
\BIBentrySTDinterwordspacing

\end{thebibliography}
\bibliographystyle{IEEEtranN}%

\end{document}